\documentclass[twocolumn,amstext,amsmath,amssymb,prd,superscriptaddress,floatfix,showpacs,nofootinbib]{revtex4}
\usepackage[dvips]{epsfig}
\usepackage{slashed}
\usepackage{mathrsfs}
\usepackage{graphicx}
\usepackage{xcolor}
\usepackage{hyperref}
\usepackage{dcolumn}
\usepackage{subfigure}
\usepackage{xspace}
\usepackage{bbm}

\definecolor{heidelbeer}{rgb}{0.5,0,0.5}

\graphicspath{
{./}
{./Figures/}
}
\begin{document}
\title{Connecting real-time properties of the massless Schwinger model to the massive case}

\author{F. Hebenstreit}
\email[]{f.hebenstreit@thphys.uni-heidelberg.de}
\affiliation{Institut f\"{u}r Theoretische Physik, Universit\"{a}t Heidelberg,
  Philosophenweg 16, 69120 Heidelberg, Germany}

\author{J. Berges}
\affiliation{Institut f\"{u}r Theoretische Physik, Universit\"{a}t Heidelberg,
  Philosophenweg 16, 69120 Heidelberg, Germany}
\affiliation{ExtreMe Matter Institute EMMI, GSI Helmholtzzentrum,
  Planckstra\ss e 1, 64291 Darmstadt, Germany}

\begin{abstract}
Quantum electrodynamics in $1 + 1$ space-time dimensions is analytically solvable for massless fermions, while no solution is known for massive fermions.
Employing the classical-statistical approach, we simulate the real-time dynamics on a lattice using Wilson fermions with mass $m$ at gauge coupling $g$.
It is shown that quantitative properties of the massless Schwinger model are emerging in the limit of large $g/m$.
We investigate two scenarios corresponding to opposite charges which are either held fixed or moving back-to-back along the light cone, as employed in effective descriptions for jet energy loss and photon production in the context of heavy-ion collisions.
Remarkably, we find that the dynamics is rather well described by the massless limit for a wide range of mass values at fixed coupling.
Moreover, our study shows that previous approximate scenarios with external charges on the light cone rather accurately capture the self-consistent dynamics of the energy conserving simulation.
\end{abstract}
\maketitle


\section{Introduction}

The real-time behavior of quantum systems composed of fermions coupled to gauge fields is one of the major challenges of modern theoretical physics.
Exactly solvable models, such as the Schwinger model of massless quantum electrodynamics (QED) in $1+1$ space-time dimensions \cite{Schwinger:1962tp}, can provide important insights into more general phenomena.
The Schwinger model shares several key properties with quantum chromodynamics (QCD) such as spontaneous chiral symmetry breaking, charge screening and the axial anomaly.
Since a long time it is used as an effective model for the description of some characteristic QCD properties~\cite{Casher:1974vf,Wong:1991ub}, and has been recently employed to real-time questions of jet energy loss and photon production in the context of ultra-relativistic heavy-ion collisions~\cite{Loshaj:2011jx,Kharzeev:2012re,Kharzeev:2013wra}. 

While the massless Schwinger model is analytically solvable, no exact solution is known for massive fermions \cite{Schwinger:1962tp,Lowenstein:1971fc,Casher:1974vf,Coleman:1975pw,Coleman:1976uz,Kiefer:1993fw,Adam:1997wt,Byrnes:2002nv,Banuls:2013jaa,Buyens:2013yza}.
The current quark masses of the light quarks are small compared to the QCD scale.
However, they are not massless and the question arises how the results of the Schwinger model are connected to the massive case.
Recently, also the prospect of constructing quantum simulators for gauge theories using ultra-cold atoms in optical lattices boosted the interest in questions regarding the real-time dynamics of string breaking in $1+1$ dimensional QED with massive fermions~\cite{Banerjee:2012pg,Hebenstreit:2013baa}. 

Though nonequilibrium real-time problems are not amenable to standard Euclidean lattice simulations, there is a large class of time-dependent problems for which the quantum dynamics can be accurately mapped onto a classical-statistical ensemble that can be simulated on a lattice~\cite{Aarts:1998td}.
This has recently been used to simulate real-time lattice QED with Wilson fermions in $1+1$ \cite{Hebenstreit:2013qxa,Hebenstreit:2013baa} as well as $3+1$ dimensions \cite{Kasper:2014uaa}.
No attempt has been made so far to connect these real-time results to the limit of vanishing fermion mass $m$.
In $1+1$ dimensions the gauge coupling $g$ has the dimension of mass and the massless limit emerges from the strongly correlated regime of large $g/m$.
Therefore, calculations in this regime serve also as an important benchmark for the simulation method where they have to connect to known exact results. 

In this work we simulate real-time properties of $1+1$ dimensional QED for a wide range of fermion mass values and demonstrate the approach to analytic predictions of the massless limit.
We investigate two scenarios corresponding to opposite charges which are either placed at a fixed distance apart from each other or moving back-to-back along the light cone.
The latter scenario has been employed for effective descriptions of jet energy loss and photon production in the context of heavy-ion collisions~\cite{Loshaj:2011jx,Kharzeev:2012re,Kharzeev:2013wra}.
Following these applications, we first consider the applied charges as external sources on the light cone.
Since the introduction of external sources violates energy conservation, the question arises whether treating the charges fully dynamically -- including the backreactions from the system onto their dynamics -- alters the results. 
For this we produce two dynamical fermions of charge $\pm g$ by applying a suitable short electric field pulse via the Schwinger effect and compare this energy conserving prescription to the previous studies. 

Remarkably, we find that for all considered cases the observed dynamics is well described by the massless limit for a large range of mass values at fixed coupling. 
This establishes an important link for applications as well as the computational procedure. 
For instance, the link clearly demonstrates the relevance of the multiple string breaking phenomenon, which was found in simulations with massive fermions \cite{Hebenstreit:2013baa}, for our understanding of the dynamics of receding fermions close to the light cone. 
Moreover, our study shows that previous approximate scenarios of external charges on the light cone rather accurately capture the self-consistent dynamics of the energy conserving simulation.     

This paper is organized as follows:
In Sec.~\ref{sec:theory} we recall some analytical results for the Schwinger model which will be needed afterwards, and we introduce the real-time lattice simulation method.
In Sec.~\ref{sec:results} we establish the link between the massive and the massless case for the scenarios of static external charges (\ref{sec:static}), receding external charges (\ref{sec:receding}), and receding dynamical charges (\ref{sec:recedingdynamical}). 
Conclusions and an outlook are given in Sec.~\ref{sec:conclusion}.

\section{QED in $1+1$ dimensions}
\label{sec:theory}

The Lagrangian density for QED in $1+1$ dimensions, which is a super-renormalizable theory, is given by
\begin{equation}
 \label{eq:lagrangian}
 \mathcal{L}=\bar{\psi}(x)[i\slashed{\partial}_x-g\slashed{A}(x)-m]\psi(x)-\frac{1}{4}F^{\mu\nu}(x)F_{\mu\nu}(x) \ ,
\end{equation}
where $g$ denotes the dimensionful gauge coupling, $m$ is the fermionic mass and $x=(x^0,x^1)\equiv(t,\bf{x})$.
The field strength tensor $F^{\mu\nu}(x)=\partial^\mu A^\nu(x)-\partial^\nu A^\mu(x)$ has only one non-trivial component $F^{10}(x)\equiv E(x)$ corresponding to the electric field. 

This model is exactly solvable for massless fermions, $m=0$ (Schwinger model) \cite{Schwinger:1962tp,Lowenstein:1971fc,Casher:1974vf}.
For massive fermions, $m\neq0$, no analytic solution is known so that one relies on approximate solutions \cite{Coleman:1975pw,Coleman:1976uz,Kiefer:1993fw,Adam:1997wt,Byrnes:2002nv,Banuls:2013jaa,Buyens:2013yza}. 
Below we employ real-time lattice gauge theory to calculate the time evolution of observable quantities such as the electric field or the fermion charge following the classical-statistical approach of Refs.~\cite{Aarts:1998td,Hebenstreit:2013qxa,Kasper:2014uaa}. 
This allows us to investigate the observables in a wide range of mass values and to approach the analytic predictions in the massless limit.

\subsection{The Schwinger model revisited}

In this section, we briefly summarize some results for the dynamics of observables which will be needed for comparison with our numerical results afterwards. 
For more detailed derivations we refer to \cite{Casher:1974vf,Iso:1988zi,Chu:2010xc}.

For massless fermions $m=0$, the theory \eqref{eq:lagrangian} can be bosonized and represented solely in terms of a real scalar field $\phi(x)$ of mass $M\equiv g/\sqrt{\pi}$, obeying the Klein-Gordon equation \cite{Lowenstein:1971fc,Casher:1974vf,Coleman:1975pw}
\begin{equation}
 \label{eq:hom_klein_gordon}
 \left[\partial_\mu\partial^\mu+M^2\right]\phi(x)=0 \ .
\end{equation}
The fermion current is related to the scalar field according to
\begin{equation}
 \label{eq:current_scalar}
 j^\mu(x)=\ :\bar{\psi}(x)\gamma^\mu\psi(x):\ =\frac{1}{\sqrt{\pi}}\epsilon^{\mu\nu}\partial_\nu\phi(x) \ ,
\end{equation}
with the two-dimensional Levi-Civita symbol $\epsilon^{\mu\nu}$.
The electric field is proportional to the scalar field:
\begin{equation}
 E(x)=\frac{g}{\sqrt{\pi}}\phi(x)=M\phi(x) \ .
\end{equation}

Coupling the Schwinger model to an external current $j^\mu_\mathrm{ext}(x)$, which we parametrize in terms of $\phi_\mathrm{ext}(x)$ in analogy to \eqref{eq:current_scalar}, gives the equation of motion 
\begin{equation}
 \label{eq:inhom_klein_gordon}
 \left[\partial_\mu\partial^\mu+M^2\right]\phi(x)=-M^2\phi_\mathrm{ext}(x) \ .
\end{equation}
As a consequence, we have for a given external charge distribution $j^{0}_\mathrm{ext}(x)$:
\begin{equation}
 \phi_\mathrm{ext}(x)=\sqrt{\pi}\int\limits^{\mathbf{x}}{d{\bf{x}'}j^0_\mathrm{ext}(t,\mathbf{x}')} \ .
\end{equation}
For suitable initial conditions, all observables can then be calculated from the solution of \eqref{eq:inhom_klein_gordon}.
In particular, the total electric field is given by
\begin{equation}
 \label{eq:electric_total}
 E(x)=\frac{g}{\sqrt{\pi}}\left[\phi(x)+\phi_\mathrm{ext}(x)\right] \ ,
\end{equation}
whereas the total charge density is
\begin{equation}
 \label{eq:charge_total}
 q(x)=\frac{1}{\sqrt{\pi}}\partial_{\bf{x}}\phi(x)+j^0_\mathrm{ext}(x) \ .
\end{equation}

\subsubsection{Static external charges}

In the first scenario, which we consider, two opposite charges of strength $\pm g$ are placed at a distance $d=2\mathbf{y}$ apart from each other at initial time $t_0=0$ \cite{Iso:1988zi,Chu:2010xc},
\begin{equation}
 j^0_\mathrm{ext}(x)=\left[\delta(\mathbf{x}+\mathbf{y})-\delta(\mathbf{x}-\mathbf{y})\right]\Theta(t) \ ,
\end{equation}
with the Dirac delta function $\delta(\mathbf{x}\pm\mathbf{y})$ and the Heaviside step function $\Theta(t)$.
As a consequence, we have
\begin{equation}
 \phi_\mathrm{ext}(x)=\sqrt{\pi}\Theta(\mathbf{y}-|\mathbf{x}|)\Theta(t) \ ,
\end{equation}
with corresponds to an initial electric field strength $g$ between the two external charges.
The solution of \eqref{eq:hom_klein_gordon} for $t\geq0$ with $\phi(t<0,\mathbf{x})=0$ is given by the integral representation
\begin{align}
 &\phi(x)=2\sqrt{\pi}M^2\times \nonumber \\
 &\int_{-\infty}^{\infty}{\frac{dk}{2\pi}\frac{\cos(k\mathbf{x})\sin(k\mathbf{y})\big[\cos(t\sqrt{k^2+M^2})-1\big]}{k(k^2+M^2)}} \ .
\end{align}
Denoting $\mathbf{x}_\pm=\mathbf{x}\pm\mathbf{y}$ and employing the Fourier cosine transformation,
\begin{align}
 \int_{0}^{\infty}{\frac{dk}{2\pi}}&\frac{\sin(t\sqrt{k^2+M^2})}{\sqrt{k^2+M^2}}\cos(k\mathbf{x}_\sigma)=\nonumber \\
 &\qquad \qquad  \quad \frac{1}{4}J_0(M\sqrt{t^2-\mathbf{x}_\sigma^2})\Theta(t-|\mathbf{x}_\sigma|)\ , 
\end{align}
the charge density can be written as
\begin{align}
 \label{eq:static_charge}
 &q(x)=\nonumber\\
 &\sum_{\sigma=\pm}\sigma\Big[\delta(\mathbf{x}_\sigma)-\frac{M^2}{2}\Theta(t-|\mathbf{x}_\sigma|)\int_{|\mathbf{x}_\sigma|}^{t}{dt'J_0(M\sqrt{t'^2-\mathbf{x}_{\sigma}^2})}\Big]
\end{align}
with $J_0$ being the zeroth order Bessel function of the first kind.
For asymptotically large times $t\to\infty$, the integrals can be performed analytically and one obtains for the total charge density \eqref{eq:charge_total}:
\begin{equation}
 \label{eq:static_screening}
 q(t\to\infty,\mathbf{x})=\sum_{\sigma=\pm}\sigma\left[\delta(\mathbf{x}_\sigma)-\frac{M}{2}e^{-M|\mathbf{x}_\sigma|}\right] \ .
\end{equation}
One recognizes that the positive/negative external charge, which has been placed at $\mp\mathbf{y}$, is exponentially screened by a cloud of negative/positive charge with the screening length $1/M$.

\subsubsection{Receding external charges}

In the second scenario, which we consider, two opposite charges of strength $\pm g$ are moving back-to-back (positive to the left, negative to the right) along the light cone after being placed on top of each other at $t_0=0$ \cite{Casher:1974vf}:
\begin{equation}
 j^0_\mathrm{ext}(x)=[\delta(\mathbf{x}+t)-\delta(\mathbf{x}-t)]\Theta(t) \ ,
\end{equation}
corresponding to
\begin{equation}
 \phi_\mathrm{ext}(x)=\sqrt{\pi}\Theta(t-|\mathbf{x}|)\Theta(t) \ .
\end{equation}
The solution of \eqref{eq:hom_klein_gordon} for $t\geq0$ with $\phi(t<0,\mathbf{x})=0$ is then given in terms of the zeroth order Bessel function of the first kind:
\begin{equation}
 \phi(x)=\sqrt{\pi}\Theta(t-|\mathbf{x}|)\left[J_0\big(M\sqrt{t^2-\mathbf{x}^2}\,\big)-1\right] \ .
\end{equation}
The electric field has strength $g$ between the two external charges.
The total electric field \eqref{eq:electric_total}, however, becomes screened such that it is given by 
\begin{equation}
 \label{eq:moving_efld}
 E(x)=g\Theta(t-|\mathbf{x}|)J_0\big(M\sqrt{t^2-\mathbf{x}^2}\,\big) \ .
\end{equation}
On the other hand, the total charge density \eqref{eq:charge_total} of the system is given in terms of the first order Bessel function of the first kind:
\begin{equation}
 \label{eq:moving_charge}
 q(x)=\frac{M\mathbf{x}}{\sqrt{t^2-\mathbf{x}^2}}\Theta(t-|\mathbf{x}|)J_1\big(M\sqrt{t^2-\mathbf{x}^2}\,\big) + j^0_\mathrm{ext}(x) \ .
\end{equation}

\subsection{Real-time lattice gauge theory}
\label{sec:method}

In this section, we briefly summarize the numerical method which is employed to investigate the real-time dynamics of model \eqref{eq:lagrangian}. 
For a detailed presentation and discussion of the range of validity of the employed classical-statistical lattice approach we refer to Refs.~\cite{Hebenstreit:2013qxa,Kasper:2014uaa}. 
Here we note that the present study represents also an important benchmark for the simulation method where it has to connect to known exact results. 

Because of the one-dimensional geometry, the gauge field dynamics is completely dictated by the fermion current such that the classical-statistical theory reduces to solving the initial value problem
\begin{subequations}
\label{eq:continuum_eom}
\begin{align}
 [i\slashed{\partial}_{x}-g\slashed{A}(x)]\Delta(x,y)&=m\Delta(x,y) \ , \\
 \partial_\mu F^{\mu\nu}(x)&=-\frac{g}{2}\operatorname{Tr}\{\Delta(x,x)\gamma^\nu\} \ ,
\end{align}
\end{subequations}
for classical gauge field configurations. Here, the Keldysh two-point function for the (quantum) fermions is defined by the commutator expectation value 
\begin{equation}
 \label{eq:keldysh}
 \Delta(x,y)\equiv\langle[\psi(x),\bar{\psi}(y)]\rangle \ .
\end{equation}

In order to solve the partial differential equation system \eqref{eq:continuum_eom}, we discretize space-time on a lattice.
We denote the spatial and temporal lattice spacing by $a_s$ and $a_t$, respectively.
Periodic boundary conditions are employed in the compactified spatial direction, $\mathbf{x}=n a_s$ with $n\in\{0,...,N-1\}$ and the total box length $L=Na_s$, whereas the time direction remains non-compact, $t=j a_t$ with $j\in\mathbb{N}^0$.

To ensure gauge invariance we use the compact formulation of $U(1)$ gauge theory by introducing the parallel transporter $U_{\mu}(x)$ which is associated with the link emanating from a lattice site $x$ and pointing in the direction of the lattice axes $\mu\in\{0,1\}$.
For simplicity, we utilize temporal-axial gauge $A_0(x)=0$, corresponding to $U_{0}(x)=1$, in our numerical simulations and we will henceforth denote
\begin{equation}
 U(x)\equiv U_{1}(x)=\exp\left(iga_sA_1(x)\right) \ .
\end{equation}

Assuming vacuum initial conditions in the fermion sector, we calculate the Keldysh two-point function \eqref{eq:keldysh} by employing a mode function expansion\footnote{Such an expansion can always be achieved since the fermions appear quadratically in the Lagrangian \eqref{eq:lagrangian} for given gauge field configuration.}
\begin{equation}
 \Delta(x,y)=\frac{1}{L}\sum_{q}[\Phi^{u}_q(x)\bar{\Phi}^{u}_q(y)-\Phi^{v}_q(x)\bar{\Phi}^{v}_q(y)] \ ,
\end{equation}
with $q\in\{-\frac{N}{2},...,\frac{N}{2}-1\}$. 
In a basis in which the Dirac gamma matrices are represented by Pauli matrices, $\gamma^0=\sigma_1$ and $\gamma^1=-i\sigma_2$, the initial values of the mode functions at $t_0=0$ are given by
\begin{subequations}
\begin{align}
 \Phi^{u}_{q}(x)&=\frac{1}{\sqrt{2\omega(\omega+p})}\begin{pmatrix}\omega+p\\\widetilde{m}\end{pmatrix}e^{2\pi i n q / N} \ , \\
 \Phi^{v}_{q}(x)&=\frac{1}{\sqrt{2\omega(\omega+p})}\begin{pmatrix}\omega+p\\-\widetilde{m}\end{pmatrix}e^{-2\pi i n q / N} \ ,
\end{align}
\end{subequations}
with
\begin{subequations}
\begin{align}
 p&=\frac{1}{a_s}\sin\left(\frac{2\pi q}{N}\right) \ , \\
 \widetilde{m}&=m+\frac{2}{a_s}\sin^2\left(\frac{\pi q}{N}\right) \ ,
\end{align}
\end{subequations}
and $\omega=\sqrt{\widetilde{m}^2+p^2}$.
Here we utilized a Wilson term to suppress the spatial doubler modes, i.e.\ to ensure that only low-momentum excitations show a low-energy dispersion relation \cite{Nielsen:1981,Wilson:1974sk}.
The mode functions then obey the lattice equation of motion
\begin{align}
 \Phi(x_{+t})=&\Phi(x_{-t})+\frac{a_t}{a_s}\gamma^0\left[U(x)(i-\gamma^1)\Phi(x_{+s})+\right.\nonumber \\
  &\left.U^*(x_{-s})(i+\gamma^1)\Phi(x_{-s})-2i(ma_s+1)\Phi(x)\right] \ ,
\end{align}
where we collectively denoted $\Phi(x)\equiv\Phi^{u,v}_q(x)$.
Moreover, we used $x_{\pm t}\equiv (t\pm a_t,\mathbf{x})$ and $x_{\pm s}\equiv(t,\mathbf{x}\pm a_s)$.
We do not include a Wilson term for the temporal doubler modes as they are naturally suppressed for suitable initial conditions and for a temporal lattice spacing being much smaller than the spatial one, $a_t\ll a_s$ \cite{Borsanyi:2008eu,Mou:2013kca,Hebenstreit:2013qxa,Kasper:2014uaa}.
For practical purposes, a ratio $a_s/a_t\simeq20$ usually suffices to guarantee that temporal doubler modes are not excited during the simulated time interval.

In the gauge sector, we introduce the electric field according to
\begin{equation}
 E(x)=\frac{1}{ga_ta_s}\operatorname{Im}[U(x_{+t})U^*(x)] \ .
\end{equation}
The equation of motion for the electric field is then given by
\begin{align}
 E(x)=E(x_{-t})-\frac{ga_t}{2}\operatorname{Re}[U(x)\operatorname{Tr}\{\Delta(x_{+s},x)(i-\gamma^1)\}] \ ,
\end{align}
where the current contribution proportional to $i$ is a consequence of the spatial Wilson term.
Moreover, the electric field is supposed to fulfill Gauss law:
\begin{align}
 E(x)=E(x_{-s})-\frac{ga_s}{2}\operatorname{Re}[\operatorname{Tr}\{\Delta(x_{+t},x)\gamma^0\}] \ .
\end{align}
This constraint has to be imposed on the initial field configuration in order to simulate in the physical subspace of the theory.
As a matter of fact, the constraint equation is conserved under the time evolution, i.e. a field configuration which fulfills Gauss law at initial times does also fulfill it at any later time.


\section{Connecting to the massless case}
\label{sec:results}

We now come to the results which are based on the real-time lattice approach presented in the previous section.
In $1+1$ dimensions the gauge coupling $g$ has the dimension of mass and in our numerical simulations we use $g$ to set the scale of our problem, measuring all quantities in units of $g$.
We introduce the notion of the critical field strength for fermion pair production, which is given by
\begin{equation}
 E_\text{cr}=\frac{m^2}{g} \ .
\end{equation}
On the other hand, the electric field generated between two external charges of strength $\pm g$ is 
\begin{equation}
 \label{eq:electric_critical}
 E_\text{stat}=g=\left(\frac{g}{m}\right)^2E_\text{cr} \ .
\end{equation}
The massless limit emerges from the regime of large $g/m$ which we denote as 'strongly correlated' since the field strength measured in units of the critical one, $E_\text{stat}/E_\text{cr}=(g/m)^2$, also becomes large.
We will consider masses in the range of $0<m\leq g$ in the following.
We note that the simulation technique is applicable for even larger masses, however, one then has to increase the strength of the two external charges to guarantee the electric field to be still in the critical regime \cite{Hebenstreit:2013baa}.

\begin{figure}[b]
 \includegraphics[width=\columnwidth]{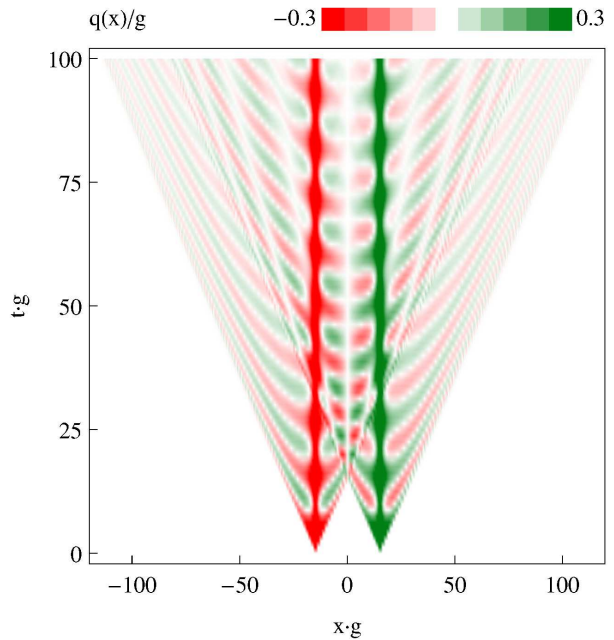}
 \caption{\label{fig:time_evolution_static} Time evolution of the charge density $q(x)$ for two opposite charges of strength $\pm g$ placed at a distance $d = 30/g$ for $g/m=20$ with lattice parameters $N=8192$ and $a_s=1/(8g)$.} 
\end{figure}

\subsection{Static external charges}
\label{sec:static}

We first consider the configuration of two opposite charges of strength $\pm g$ which are placed a distance $d=2\mathbf{y}$ apart from each other. 
Fig.~\ref{fig:time_evolution_static} shows the time evolution of the charge density for $\mathbf{y}=15/g$ and $g/m=20$. 
At early times, one observes that the induced charge density emanates from the external sources in a causal way, i.e.\ within the light cone. 
Moreover, there is a complicated interplay between the charges created from the two distinct sources which starts overlapping. 
In fact, this results in an oscillatory pattern which completely disappears at asymptotic times. 
Effectively, the asymptotic charge density around the external charges is reached after about $t \sim 400/g$.

\begin{figure}[t]
 \includegraphics[width=\columnwidth]{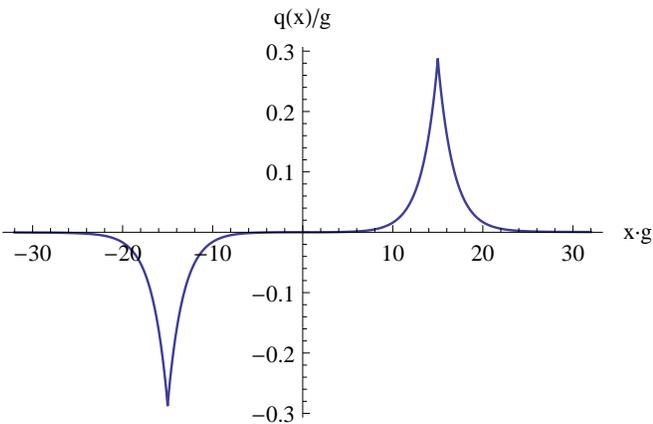}
 \caption{\label{fig:asymptotic_static} The charge density distribution around the opposite external charges at time $t=400/g$ with parameters as in Fig.~\ref{fig:time_evolution_static}.} 
\end{figure}

The complete screening of the external charges by a cloud of opposite charge at asymptotically large times is one of the most striking feature of this configuration. 
The charge density at $t \sim 400/g$ is shown in Fig.~\ref{fig:asymptotic_static}, where the absence of the earlier oscillatory pattern is visible.
For massless fermions, a similar behavior can be reproduced by integrating the two Bessel functions \eqref{eq:static_charge}.
We find that its asymptotic form \eqref{eq:static_screening} agrees rather well with the simulation results for the employed non-zero mass value.

To quantify this assertion, we investigate the asymptotic charge density in more detail as a function of mass $m$. 
To this end, we assume that the screening cloud around the external charge $-g$ at location $\mathbf{y}$ takes the functional form
\begin{equation}
 \label{eq:fitting}
 q_{\text{scr},\mathbf{y}}(\mathbf{x})=A\exp\left(-\frac{|\mathbf{x}-\mathbf{y}|}{\lambda}\right) \ ,
\end{equation}
where we use the amplitude $A$ and the screening length $\lambda$ as our fit parameters.
In the Schwinger model with massless fermions we have the asymptotic result $A=M/2$ and $\lambda=1/M$ according to \eqref{eq:static_screening}.

\begin{figure}[b]
 \includegraphics[width=0.95\columnwidth]{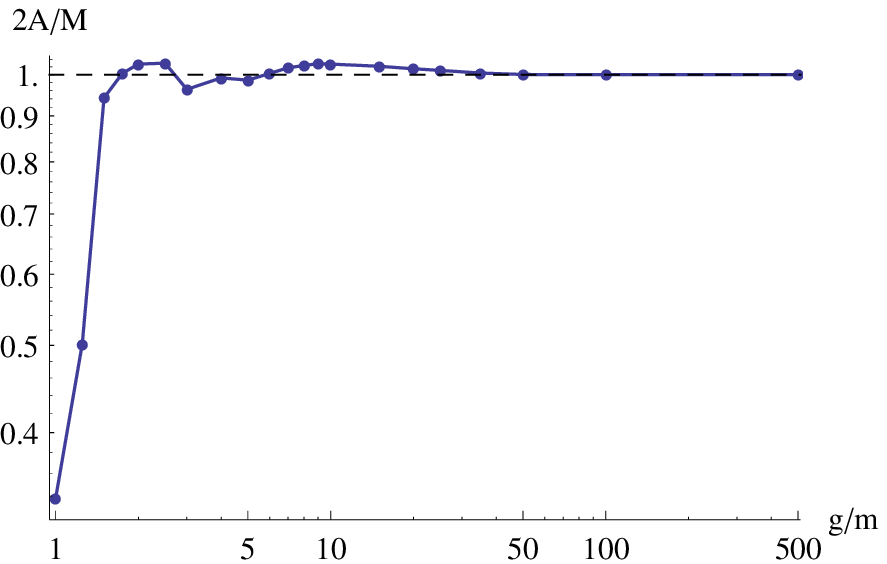}
 \includegraphics[width=0.95\columnwidth]{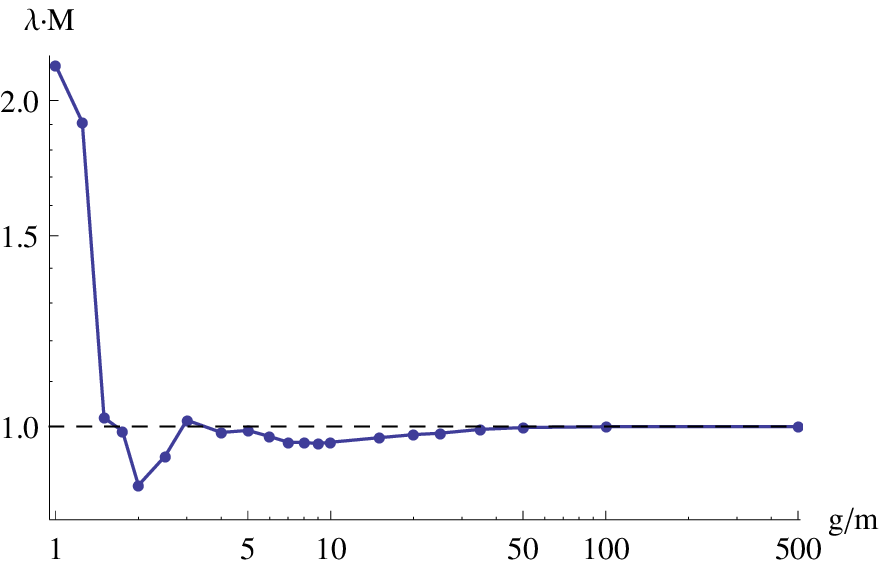} 
 \caption{\label{fig:numerical_fit} Numerical fit of $A$ ({\it upper panel}) and $\lambda$ ({\it lower panel}) in a wide range of mass values with $N=8192$, $a_s=1/(8g)$. 
 The dashed line indicates the analytical value for $m=0$.}
 \end{figure}

Fig.~\ref{fig:numerical_fit} shows the numerical fit of $\{A,\lambda\}$ for a wide range of mass values along with a comparison to the massless analytic results. 
We note that when doing the numerical fit we also perform a time average to get rid of the small oscillations which are still present at finite times. 
One observes that the results from our numerical simulations accurately connect for small enough masses to the results of the Schwinger model.
It also highlights the accuracy of the lattice method in this strongly correlated regime with large $g/m$ of order hundreds. 

This is particularly remarkable for the employed Wilson fermions, which require very large lattices to describe the physics of massless fermions.
We note that high-momentum modes are affected by the Wilson term described in Sec.~\ref{sec:method}, which vanishes in the naive continuum limit $a_s\to 0$. 
We conclude that the chosen lattice spacing $a_s$ is already close to the continuum limit. 
This is also supported by a careful analysis for an even smaller value $a_s=1/(16g)$, which showed only small differences compared to the simulation with $a_s=1/(8g)$.
We also checked that the results are insensitive to increasing the lattice size from $N=4096$ to $N=8192$.
We emphasize that we precisely approach the considered massless continuum results in our non-equilibrium study as the mass parameter $m$ is taken to be sufficiently small.
This is in contrast to Euclidean lattice simulations with Wilson fermions which require in general an additive mass renormalization to approach the chiral limit \cite{Gattringer:1999gt}.

\begin{figure*}[t]
 \includegraphics[width=\columnwidth]{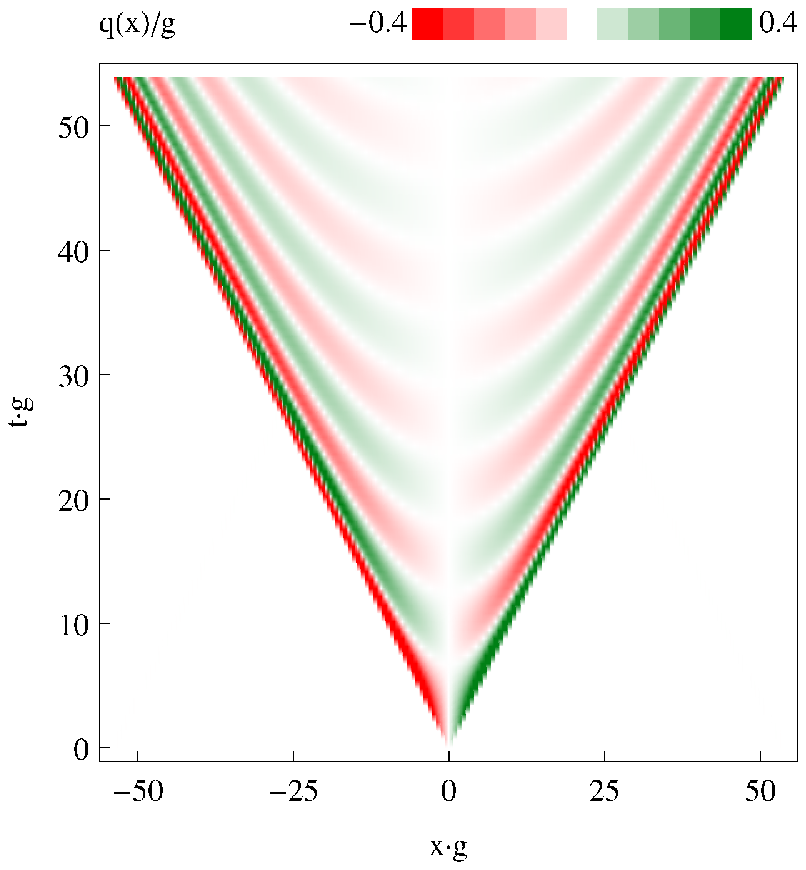}
 \includegraphics[width=\columnwidth]{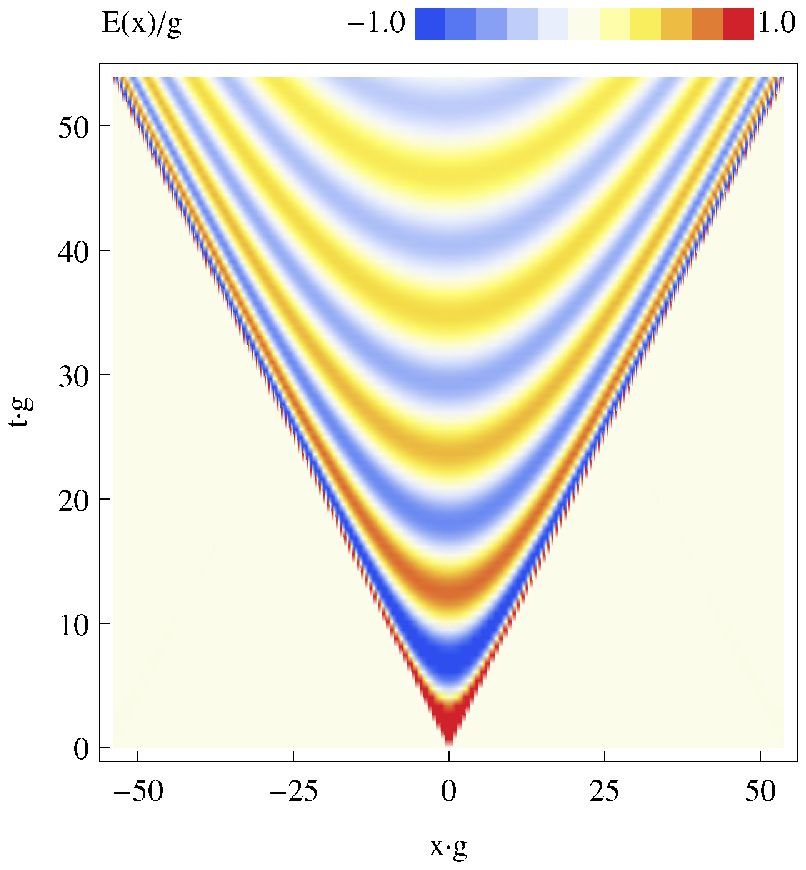} 
 \caption{ \label{fig:time_evolution_dynamic} Time evolution of the charge density $q(x)$ ({\it left panel}) and the electric field $E(x)$ ({\it right panel}) for two opposite charges of strength $\pm g$ moving back-to-back along the light cone with $g/m=100$ for lattice parameters $N=8192$ and $a_s=1/(75g)$.}
\end{figure*}
 
Increasing the mass such that $g/m\gtrsim10$, we observe a monotonic behavior of the fit parameters $\{A,\lambda\}$.
The numerical data in this regime suggests a power series behavior: 
\begin{subequations}
\begin{align}
 A&=\frac{M}{2}\left[1+A_1\left(\frac{m}{g}\right)+A_2\left(\frac{m}{g}\right)^2+...\right] \ , \\
 \lambda&=\frac{1}{M}\left[1+\lambda_1\left(\frac{m}{g}\right)+\lambda_2\left(\frac{m}{g}\right)^2+...\right] \ .
\end{align}
\end{subequations}
According to the available numerical data, we have $A_1\simeq-\lambda_1\simeq0.5$ and $A_2\simeq-\lambda_2\simeq2$.
A more precise determination of the subleading Taylor coefficients is beyond the scope of the current study as it would require substantial computational resources on the required large lattices.

For even larger mass values in the range $10\gtrsim g/m\gtrsim2$ the behavior of the fit parameters $\{A,\lambda\}$ becomes non-monotonic. 
Nevertheless, the deviation of the fit parameters $\{A,\lambda\}$ from their analytic values for $m=0$ is only of the order of a few percent. 
In addition, the functional form of the charge density at late times is still extremely well described by \eqref{eq:fitting}. 
This can be concluded from the small standard error in the numerical fits, where the error bars are smaller than the symbol sizes of the data points employed in Fig.~\ref{fig:numerical_fit}.

The behavior, however, changes drastically if we approach $g/m\to1$.
In this regime the fit parameters $\{A,\lambda\}$ start to deviate considerably from their analytical values for $m=0$.
We note that drastic changes are expected around $g/m\simeq1$ due to the transition from the overcritical regime to the subcritical regime.

In summary, the analysis of the asymptotic charge density reveals a remarkably accurate agreement in a wide range of mass values with the analytical estimates based on the massless Schwinger model.
So far, the detailed analysis above considered an asymptotic quantity. 
The specific behavior of the quantity at earlier times, on the other hand, may still show more pronounced deviations and we will come to this point in the next section.

\subsection{Receding external charges}
\label{sec:receding}

In the following, we consider two opposite charges of strength $\pm g$ which are moving back-to-back along the light cone after being placed on top of each other. 
This configuration has been recently employed in Refs.~\cite{Loshaj:2011jx,Kharzeev:2012re,Kharzeev:2013wra}. 
Such a scenario with external sources on the light cone only approximately holds for a highly relativistic configuration. 
The approximation has the advantage that the analytical results \eqref{eq:moving_efld} and \eqref{eq:moving_charge} take particularly simple forms. 
However, energy conservation is not fulfilled for this configuration as a consequence of the externally guided sources. 
We will discuss the viability of this approach in detail in the next section by comparing it to the behavior of receding dynamical charges, where energy is conserved. 

For the receding external charges the time evolution of the electric field and the charge density for $g/m=100$ is shown in Fig.~\ref{fig:time_evolution_dynamic}. 
We note that for the given parameter set the results are already very close to the massless limit as described by \eqref{eq:moving_efld} and \eqref{eq:moving_charge}. 
The external charges initially produce a constant electric field between them which then decays by the production of fermions (primary production). 
These newly created charges are themselves accelerated apart from each other such that they try to screen the external ones. 
Due to the fact that the electric field evolution is governed by the fermion current, a secondary string between the newly created charges is formed which again decays because of screening by fermions (secondary production). 
This behavior continues, though it becomes less and less efficient with time. 
Primary and subsequent secondary productions of fermions have been discussed in detail in Ref.~\cite{Hebenstreit:2013baa} in the context of the phenomenon of multiple string breaking.

\begin{figure}[t]
 \includegraphics[width=\columnwidth]{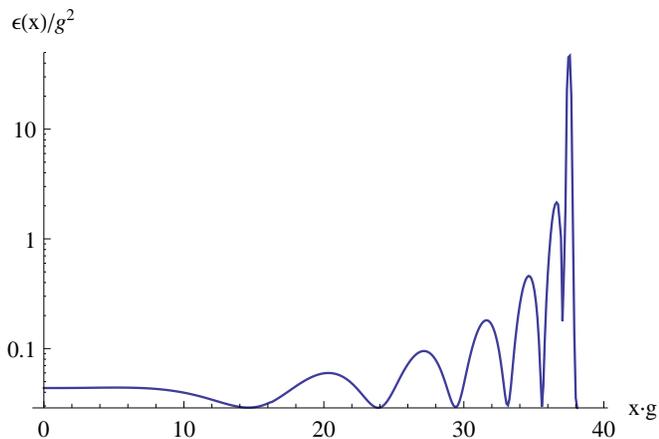}
 \caption{\label{fig:energy_density_dynamic} Fermion energy density $\epsilon(t\simeq38/g,\mathbf{x})$ on a logarithmic scale, with $\epsilon(t,-\mathbf{x})=\epsilon(t,\mathbf{x})$. 
 The parameters are as in Fig.~\ref{fig:time_evolution_dynamic}.} 
\end{figure}

We additionally display the fermion energy density at $t=38/g$ in Fig.~\ref{fig:energy_density_dynamic}.
One observes that most of the fermionic energy is contained in the primary bunch which is concentrated at the external charge whereas the subsequently produced charges do not carry much energy.
This shows that the primary fermion production is most efficient as the electric field is maximal then, whereas the production events in the subsequently formed, weaker electric strings are less pronounced.
 
\begin{figure}[b]
 \includegraphics[width=\columnwidth]{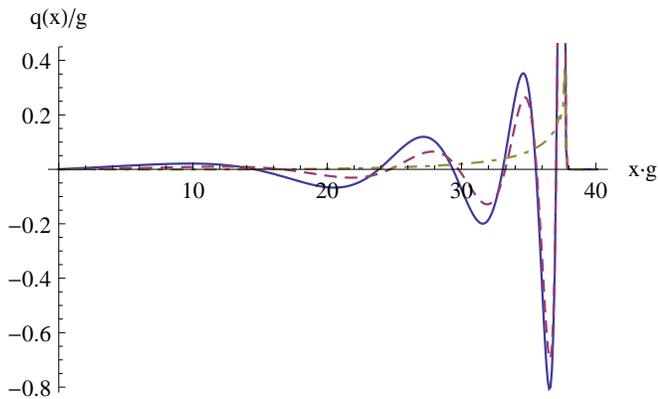}
 \caption{\label{fig:different_masses_dynamic} Charge density $q(t\simeq 38/g,\mathbf{x})$ for $g/m=100$ (solid line), $g/m=10$ (dashed line) and $g/m=1$ (dot-dashed line), with $q(t,-\mathbf{x})=-q(t,\mathbf{x})$.
 The lattice parameters are $N=8192$ and $a_s=1/(75g)$.} 
\end{figure}

Fig.~\ref{fig:different_masses_dynamic} shows the charge density at $t=38/g$ for different values of $g/m$.
Similarly to the previous discussion for the static external charges, the analytic result for massless fermions are reproduced already remarkably well for $g/m\simeq100$.
For smaller values of $g/m\simeq10$, one observes that the characteristics of the charge density remains similar to the massless case, however, the behavior changes on a quantitative level:
The overall amplitude of the charge density decreases, which can be traced back to the fact that fermion production via the Schwinger mechanism becomes less and less efficient for increasing mass values.
We still observe an oscillatory behavior of the charge density which is a consequence of the multiple fermion production events.
However, for $g/m\simeq1$ the qualitative behavior differs substantially as the charge density does not show an oscillatory pattern anymore. 
The reason is that the electric field produced by two opposite charges of strength $\pm g$ is exactly critical for $g/m=1$ according to \eqref{eq:electric_critical}.
As a consequence, fermion production becomes exponentially suppressed such that no secondary electric string is created.

\subsection{Receding dynamical charges}
\label{sec:recedingdynamical}

Finally, we go beyond the configuration of external charges by considering the two oppositely moving charges as being part of the system.
To this end, we produce two total charges $\pm g$ by applying a very short external electric field pulse via the Schwinger effect \cite{Hebenstreit:2011wk,Hebenstreit:2013qxa}. 
As a consequence, the quantum expectation values for the fermions are characterized by a finite spatial and momentum width as compared to point-like external charges moving on the light cone. 
Moreover, energy conservation is fulfilled during the subsequent time evolution.

\begin{figure}[b]
 \includegraphics[width=\columnwidth]{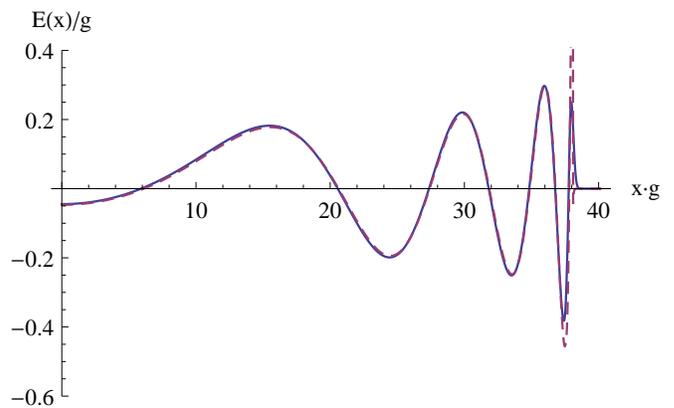}
 \caption{\label{fig:comp_external_ur} Electric field $E(t\simeq38/g,\mathbf{x})$ for self-consistent fermions (solid line) and external charges (dashed line) for $g/m=100$, with $E(t,-\mathbf{x})=E(t,\mathbf{x})$.
 The lattice parameters are $N=8192$ and $a_s=1/(75g)$.} 
\end{figure}

We first consider two receding, nearly massless fermions with $g/m=100$, initial peak momentum $p/m\simeq700$ and spatial width $\Delta x\simeq2/(5g)$.
In Fig.~\ref{fig:comp_external_ur}, the electric field of this configuration is compared to the electric field created by two external charges moving along the light cone at $t=38/g$.
The remaining numerical parameters are $N=8192$ and $a_s=1/(75g)$.

For this ultrarelativistic configuration, we observe only minor differences in the vicinity of the initial charges whereas the behavior in the space between them remains unaffected.
Moreover, we observe good agreement with the analytic expression \eqref{eq:moving_efld} for the chosen parameters.
According to the previous discussion, we know that the initial charges (both external ones as well as self-consistent ones) are screened very rapidly in the primary fermion production event.
As a consequence, a charge neutral compound is formed whose detailed charge distribution depends on whether an external or a self-consistent charge is considered. 
In contrast, the subsequent dynamics in the space between the newly formed compounds is almost independent of these details.
Moreover, the approximately charge neutral compounds do not interact with the subsequently produced charges such that they approximately retain their energy.
As a consequence, pulling the external charges further apart practically does not cost any additional electric field energy as they are already screened.
Therefore, we conclude that the dynamics of the Schwinger model for two receding fermions in the ultrarelativistic limit is remarkably well described by two external charges moving back-to-back along the light cone.

\begin{figure}[t]
 \includegraphics[width=\columnwidth]{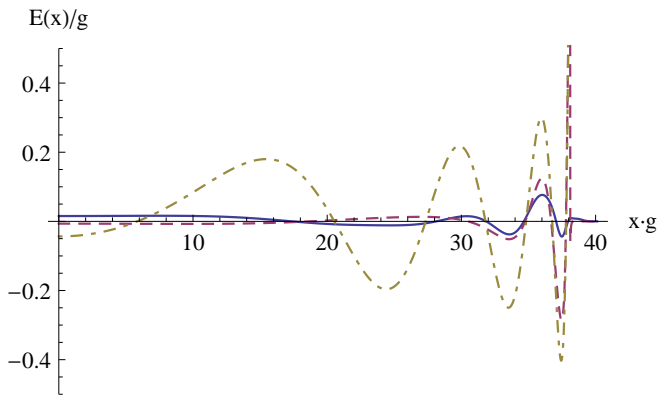}
 \caption{\label{fig:comp_external_r} Electric field $E(t\simeq38/g,\mathbf{x})$ for self-consistent fermions (solid line) and external charges (dashed line) for $g/m=5$, as well as the analytic result for massless fermions \eqref{eq:moving_efld} (dot-dashed line).
 The lattice parameters are $N=8192$ and $a_s=1/(75g)$.} 
\end{figure}

This observation may be explained further by means of a simple estimate.
To this end we consider the initial fermionic energy, which is given by
\begin{equation}
 \mathcal{E}_\text{ferm}=2\sqrt{m^2+p^2} \ .
\end{equation}
On the other hand, the energy of the electric field string, which is formed between the two receding charges, reads
\begin{equation}
 \mathcal{E}_\text{el}=\frac{L_1g^2}{2} , 
\end{equation}
where $L_1$ denotes the distance at which the primary fermion production occurs.
According to the results of Fig.~\ref{fig:time_evolution_dynamic} we have $L_1g\simeq5$.
Requiring that the fermion energy loss must be smaller than the electric field energy, $\mathcal{E}_\text{ferm}>\mathcal{E}_\text{el}$, we find the following condition on the initial fermion momentum:
\begin{equation}
 \label{eq:estimate}
 \frac{p}{m}>\frac{g}{m}\sqrt{\left(\frac{L_1g}{4}\right)^2-\left(\frac{m}{g}\right)^2} \ .
\end{equation}
This condition is clearly fulfilled by a factor of $5.5$ for $g/m=100$ and the employed initial fermion momentum $p/m\simeq700$.

Choosing a configuration of relativistic, massive fermions with only $g/m=5$, the behavior may change considerably.
In the following, we consider a configuration with initial peak momentum $p/m\simeq9.5$ and spatial width $\Delta x\simeq 1/g$. 
We note that the estimate \eqref{eq:estimate} is only fulfilled by a factor of $1.5$ in this case. 
In Fig.~\ref{fig:comp_external_r}, we again compare the resulting electric field with the one created by two external charges, as well as the analytic prediction for massless fermions \eqref{eq:moving_efld}.

We observe two major discrepancies which have different origins:
On the one hand, comparing the curves corresponding to the external charges and the analytical prediction \eqref{eq:moving_efld}, we again encounter the effect of an explicit fermion mass as discussed in the previous section for these values of $g/m$. 
Additionally, as compared to the ultra-relativistic case, we observe differences between the self-consistent and the external charge configuration.
The distinctions, however, concern only the very details of the electric field but not the overall amplitude and the general oscillatory behavior.
The self-consistent dynamics of massive fermions in $1+1$ dimensions with relativistic energies is still remarkably well-described by two external charges moving back-to-back along the light cone.


\section{Conclusion \& outlook}
\label{sec:conclusion}

We investigated the real-time dynamics QED in $1+1$ dimensions using real-time lattice simulations with Wilson fermions in the classical-statistical approach. 
As initial conditions we chose two opposite charges which were either held fixed or were moving back-to-back along the light cone. 
For $g/m>1$, these charges create an overcritical electric field of strength $g$ according to Gauss law in $1+1$ dimensions, which then decays via the Schwinger mechanism.

Regarding the configuration of two external charges which are held fixed, we analyzed the resulting asymptotic charge distribution in detail.
We found that the charge distribution, whose exponential form is analytically known for massless fermions, is very robust in a wide range of finite mass values $g/m>1$.
Most notably, for $g/m\gtrsim100$ we found very good agreement of our numerical results with the analytic ones. 
This represents an important benchmark for the simulation method, also in view of the employed Wilson fermions in the small mass regime. 
For $1\lesssim g/m\lesssim100$, the charge density still exhibits an exponential form, however, the corresponding parameters start deviating from the massless case.
Only for $g/m\to1$ substantial deviations were found, which were attributed to the transition from the overcritical to the subcritical regime.

We then studied the dynamics of two external charges moving back-to-back along the light cone by investigating the charge density and the electric field.
Again, we found excellent agreement between our numerical results for $g/m\gtrsim100$ and the analytical predictions for massless fermions. 
For increasing masses $g/m\to1$, however, the behavior gradually changes since fermion production via the Schwinger mechanism becomes less and less efficient.
Finally, considering the receding charges as being part of the system, we found that the dynamics of fermions with relativistic energies is still well-described by two external charges moving back-to-back along the light cone.

In view of the fact that QED in $1+1$ dimensions for massless fermions has been frequently employed as an effective model for characteristic aspects of QCD, our results provide two important insights:
First, it seems to be well-justified to neglect fermion masses for $g/m>1$ to a great extent.
More importantly, however, our results also suggest that describing the receding charges as external sources is a good approximation even though this description violates energy conservation. 
The reason is that the initial charges are anyhow screened very rapidly by the primary fermion production event.
As a consequence, the newly formed compounds do not interact with the subsequently produced charges and pulling the external charges further apart practically does not cost any additional electric field energy.


\subsection*{Acknowledgments}
  
We thank O.~Islam for many discussions on the relation between the massless Schwinger model and the massive case.
We thank D.~Gelfand and V.~Kasper for discussions and collaborations on related work.
F.~Hebenstreit is supported by the Alexander von Humboldt Foundation.


\end{document}